# Distributions of Charged Hadrons Associated with High $p_T$ Particles

**Fuqiang Wang (for the STAR Collaboration)**

Department of Physics, Purdue University, West Lafayette, Indiana 47907, USA

fqwang@purdue.edu

**Abstract**. Recent results from STAR on angular correlations of charged hadrons associated with high transverse momentum particles are presented. Emphasis is put on the correlations of soft associated hadrons, which provide rich information on the properties of the bulk medium created in RHIC collisions. The results are discussed in the context of three pertinent questions regarding QGP formation: sufficient energy density, parton thermalization, and deconfinement.

Quantum Chromodynamics (QCD) predicts a phase transition between hadronic matter and quark-gluon plasma (QGP), a *deconfined* and *thermalized* state of quarks and gluons, at a critical energy density of approximately 1 GeV/fm$^3$ [1]. Such a phase transition is being actively pursued at the Relativistic Heavy Ion Collider (RHIC) at BNL via ultra-relativistic heavy-ion collisions.

The RHIC experiments [2] have critically assessed the three-year wealth of experimental data and concluded that a new form of hot and dense matter has been created in RHIC collisions. Theory and theoretical models, confronting available experimental data, conclude that the matter created in RHIC collisions is a strongly coupled QGP [3]. In this proceedings article we focus on angular correlations with high transverse momentum particles and try to address three pertinent questions regarding QGP formation:

(1) Is the energy density created in RHIC collisions sufficiently high for QGP formation?
(2) Is the created system a thermalized state of quarks and gluons?
(3) Are the quarks and gluons deconfined over a volume of nuclear size?

## 1. Energy Density

In the Bjorken boost invariant picture [4], initial energy density can be estimated from the produced total transverse energy. Such estimates indicate an energy density of 4-5 GeV/fm$^3$ at proper time of 1 fm/c for central Au+Au collisions at RHIC [5]. Such large energy densities (which are also estimated at the CERN SPS [6]), taken at face values, are well above the predicted critical energy density for phase transition [1].

A perhaps more "direct" way to gauge initial energy density is to "send" a self-generated probe – hard-scattering partons – through the created medium and measure any modifications to the probe by the medium. Due to the large bombarding energies at RHIC, high transverse momentum ($p_T$) particles become statistically abundant for the first time in heavy-ion collisions. High $p_T$ particles come predominantly from jets emerging from initial hard-scatterings between partons. They need finite time to escape the collision zone, during which a dense medium is formed. The partons and fragmented hadrons are expected to lose energy via interactions with the medium, resulting in the so-called jet quenching phenomena - suppression of inclusive yield and angular correlation strength at high $p_T$ [7].

The larger the medium gluon density is, the stronger the interaction and the larger the suppression magnitude. Thus, high $p_T$ particles and jet quenching provide a powerful, direct tool to measure the medium density created in ultra-relativistic heavy-ion collisions.

1.1. High $p_T$ Suppression

All four RHIC experiments have observed suppressed yields at high pT in central Au+Au collisions per binary nucleon-nucleon collision [8,9,10,11]. Figure 1 (from [12]) shows the centrality dependence of the single inclusive hadron suppression factor, $R_{AA}$, for charged hadrons at $p_T > 6$ GeV/c measured by the STAR collaboration [9] and for neutral pions measured by the PHENIX collaboration [8]. The suppression rate increases gradually with centrality, up to about 4 in the most central 5% collisions.

For the observed high $p_T$ particles, whose overall yield is strongly suppressed, STAR has measured the correlated hadrons in azimuthal angle in the $p_T$ range of $2 < p_T < 4$ GeV/c [13,14]. While the jet-like angular correlation strengths are similar on the near side between pp and central Au+Au, those on the away side are strongly suppressed at large $p_T$ [13,14].

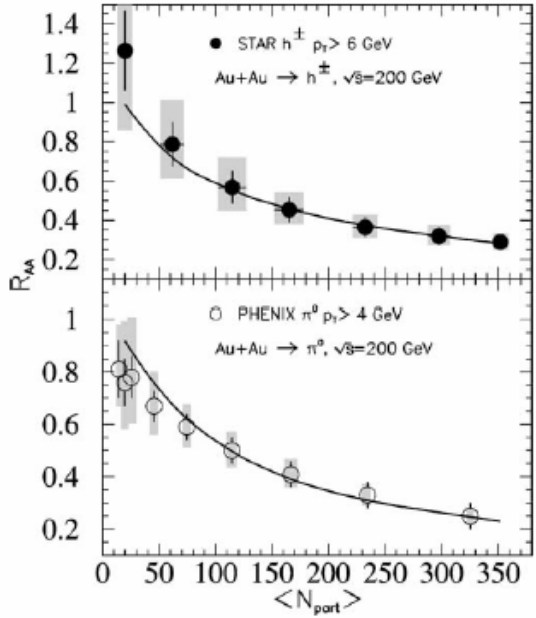

**Figure 1.** The centrality dependence of the measured single inclusive hadron suppression factor at high $p_T$ [8,9] as compared to theoretical calculation with parton energy loss [12]. The figure is taken from Ref. [12].

1.2. Inferred Energy Density

The observed high-$p_T$ suppressions could in principle come from, not the final state jet quenching effect discussed above, but initial state effects present in the Au nuclei (such as saturation of gluon density [15]). In order to discriminate between initial and final state effects, d+Au collisions were measured by all four RHIC experiments. The measured d+Au results [10,16] are similar to the pp results at high $p_T$, demonstrating that the suppression phenomena in central Au+Au collisions are due to final state interactions. Invoking partonic energy loss due to final state interactions of hard partons with the created medium, and assuming that all high $p_T$ hadrons coming from parton fragmentation, perturbative QCD (pQCD) model calculations [17] indicate a gluon density of about 30 times (or energy density of about 100 times) that of normal nuclear matter in order to reproduce the observed magnitudes of high $p_T$ suppression. The curves in Fig. 1 show such a calculation. Note that the inferred energy density from pQCD model calculations is in general agreement with the Bjorken energy density estimate that is from a completely different line of argument. The agreement reassures that a high enough energy density may indeed have been achieved in central Au+Au collisions for the predicted phase transition.

## 2. Thermalization

The QGP is a state of thermalized quarks and gluons. If thermalization is reached at an initial stage, the measured final stage hadrons will possess thermal distributions. Indeed, soft physics results on particle spectra and yields [18] and event-by-event $<p_T>$ fluctuations [19] indicate a chemically and (local-)kinetically equilibrated system at the final freeze-out stage. This is of course a necessary, but not sufficient condition for early thermalization. Meanwhile elliptic flow ($v_2$), established early in the collision because of its self-quenching nature, is found to be well described at low $p_T$ by ideal hydrodynamic calculations [20]. Since ideal hydrodynamic fluid is a thermalized system with zero mean-free-path, they give the maximum possible $v_2$ value. Thus, the consistency between the measured $v_2$ and hydrodynamics result suggests an early thermalization in heavy-ion collisions at RHIC.

Another, perhaps more "direct" way to investigate thermalization is to study the energy distribution of final hadrons related to initial hard-scattered partons after interactions with the medium by the partons, hadrons, or both. In other words, we have effectively two distinct sources of particles: jet fragments, which are initially hard (or would be in the absence of a medium), and decay products of the bulk medium, which are generally soft. Particles from the two sources are brought together to interact and, as an inevitable consequence, reach equilibration with each other.

### 2.1. Hard-Soft Angular Correlations

The away side associated hadrons (opposite to a high $p_T$ trigger particle) are significantly depleted at large $p_T$ [13,14]. The depleted energy must be redistributed to low $p_T$ particles. Reconstruction of these low $p_T$ particles will serve as an experimental confirmation of jet quenching. But moreover, complete reconstruction of jets and study of their modifications will provide information about the properties of the nuclear medium [12,21]. For example, by studying the amount of energy loss and how the energy is distributed, one may experimentally learn about the medium density, the underlying energy loss mechanism(s), and the degree of equilibration between the energy and the medium.

STAR has performed a nearly complete statistical reconstruction of charged hadron jets by correlating charged hadrons at all $p_T$ with a leading particle and subtracting the combinatoric background [14]. The background is obtained from mixed-events, with the elliptic flow modulation added in by hand and normalization fixed to the correlation function in the $0.8 < |\Delta\phi| < 1.2$ region [14]. Figure 2 shows the background subtracted angular correlation functions in pp and the 5% most central Au+Au collisions with trigger particle transverse momentum of $4 < p_T^{trig} < 6$ GeV/c and associated particle $0.15 < p_T < 4$ GeV/c [14]. The systematic uncertainties mainly come from uncertainties in background normalization and in the measured elliptic flow [14]. In contrast to correlations at high $p_T$, it is found that the low $p_T$ hadrons are present on the away side of the trigger particle (away side being $|\Delta\phi| > 1.0$ and $|\eta| < 1.0$) and there are more of them in central Au+Au collisions than in pp. It is also found that the $\Delta\phi$ correlation becomes broader on the away side in central Au+Au collisions, reaching the limit that can be described [14,22] by the so-called statistical momentum balance [23]. The away-side associated particles, in fact, no longer appear jet-like because large $p_T$ particles are essentially depleted. Instead, what are left on the away side, in the picture of initial di-jet production, are the remnants of the away-side jet after intensive interactions with the medium.

On the near side ($|\Delta\phi| < 1$ and $|\Delta\eta| < 1.4$), correlation in $\Delta\phi$ remains similar, while that in $\Delta\eta$ becomes broader in central Au+Au collisions; The $\Delta\eta$ correlation in central Au+Au collisions appears to have two components: a Gaussian peak on a more or less flat pedestal. Such a two-component feature is not present in pp and d+Au collisions [24]. One may separate the correlated yield into two parts: one contained in the Gaussian peak and the other in the pedestal. The Gaussian peak yield is found to be rather invariant between pp, d+Au, and Au+Au collisions at all centralities. The Gaussian peak width is broader in central Au+Au than in pp or d+Au for $4 < p_T^{trig} < 6$ GeV/c [24]. The underlying reason(s) for the broad $\Delta\eta$ distribution or the pedestal-Gaussian structure are not yet identified; there could be a number of reasons, such as longitudinal [25] and transverse [26] flow developed in the medium.

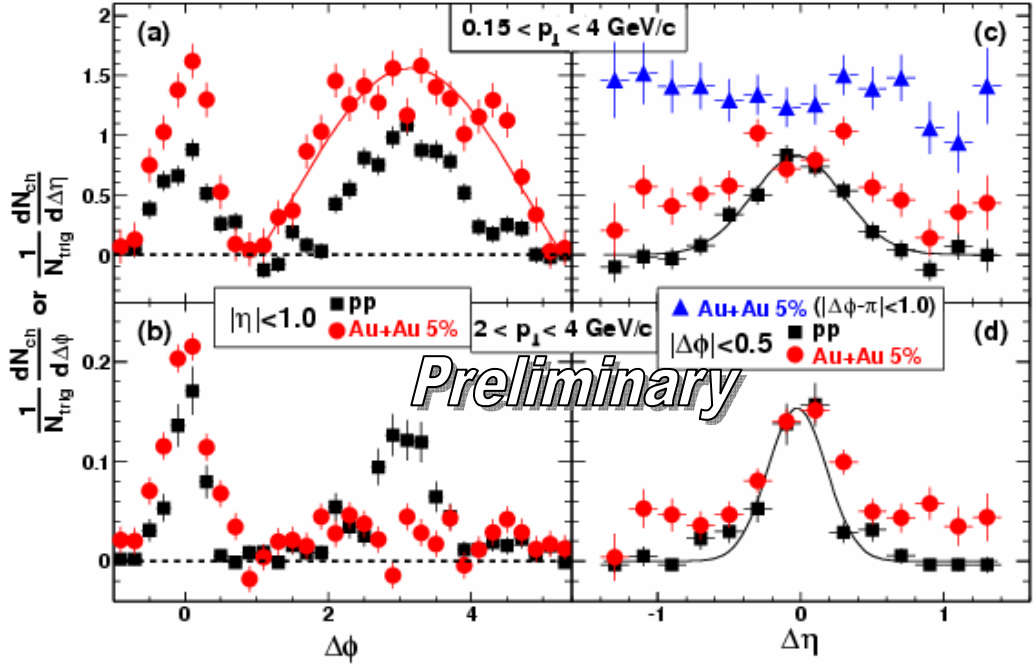

**Figure 2.** Background subtracted, per trigger particle normalized (a,b) $\Delta\phi$ and (c,d) $\Delta\eta$ correlation functions for pp and the 5% most central Au+Au collisions between trigger particles of $4 < p_T^{trig} < 6$ GeV/c and associated hadrons of (a,c) $0.15 < p_T < 4$ GeV/c and (b,d) $2 < p_T < 4$ GeV/c [14]. The subtracted background level for associated $p_T$=0.15-4 GeV/c is approximately 1.4 in pp and 211 in the 5% most central Au+Au; That for associated $p_T$=2-4 GeV/c is approximately 0.007 in pp and 2.1 in the 5% most central Au+Au.

2.2. Away-Side Associated $<p_T>$

Figure 3 shows the $p_T$ spectra of the associated hadrons on the near side and away side. While the near-side hadron spectral shapes are similar in pp and Au+Au, the away-side associated hadrons are significantly softened in central Au+Au collisions, becoming similar to those from the bulk [14]. The lost energy at high $p_T$ has been moved to low $p_T$. By triggering on a high $p_T$ particle in the final state, due to significant energy loss, only those di-jets produced near the surface of the medium are selected. The two jet partners encounter quite different amounts of medium: The near-side jet goes through a thin layer of medium, losing a modest amount of energy, and emerges with a leading particle at high $p_T$ accompanied by hadrons with a minimally modified $p_T$ distribution; The away-side jet goes deep into the medium, suffering significant energy loss, and emerges with a collective excess of particles with a significantly softened $p_T$ distribution and with little jet-like characteristics surviving. Recent calculations of large angle hadron correlations from medium-induced gluon radiation can qualitatively describe our results [27].

Figure 4 shows the $<p_T>$ of the away-side associated hadrons and of inclusive hadrons in p+p and as a function of centrality in Au+Au collisions from STAR [14]. The $<p_T>$ of associated hadrons decreases with centrality, while that of inclusive hadrons increases with centrality due to collective radial flow. In central collisions the $<p_T>$ of associated hadrons is not much larger than that of inclusive hadrons. Particles from the two distinct sources, one from initial hard-scattered partons (or jets) and the other from the bulk medium, appear to have reached partial but significant degree of thermalization due to intensive interactions on the away side. Hard-soft parton interactions are likely responsible for the observed partial thermalization. Because soft-soft parton interactions within the medium are much stronger than hard-soft parton interactions, the observed softening in Fig. 4 may in turn indicate a high degree of *thermalization* in the medium itself.

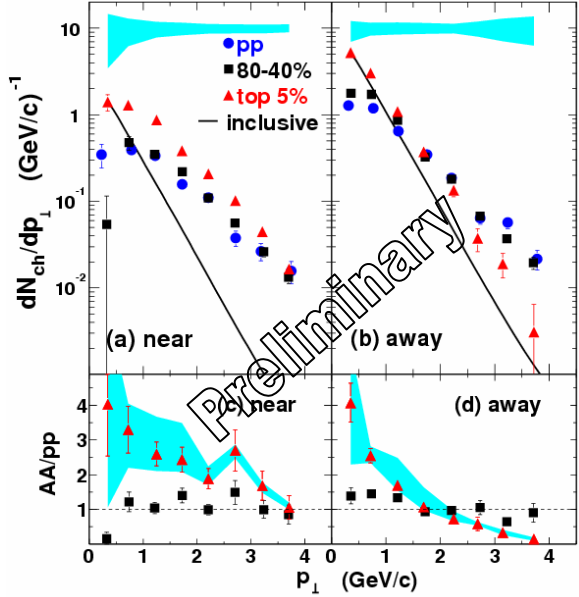 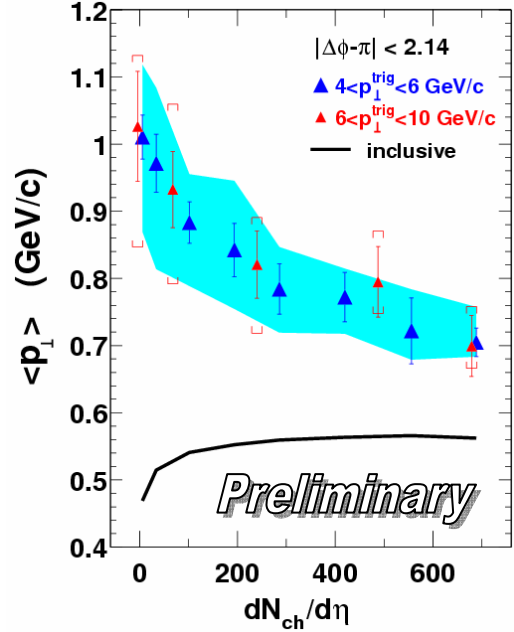

**Figure 3.** Near-side (a) and away-side (b) $p_T$ distributions of associated charged hadrons with trigger particle $4 < p_T^{trig} < 6$ GeV/c in pp, peripheral and central Au+Au collisions, and ratios of Au+Au to pp distributions for near-side (c) and away-side (d) [14]. Errors shown are statistical. The bands show the systematic errors for central collisions. The lines are $p_T$ distributions of inclusive hadrons in central collisions.

**Figure 4.** Associated charged hadron $\langle p_T \rangle$ within $0.15 < p_T < 4$ GeV/c on the away side in pp (the leftmost set of data points) and Au+Au collisions [14]. The trigger particle $p_T$ ranges are $4 < p_T < 6$ GeV/c (large triangles, systematic uncertainties shown by the shaded area) and $6 < p_T < 10$ GeV/c (small triangles, systematic uncertainties shown by the caps). The line shows $\langle p_T \rangle$ of inclusive hadrons.

2.3. Away-Side Associated $\langle p_T \rangle$ versus $\Delta\phi$

Due to the collision geometry, the associated $\langle p_T \rangle$ may depend on the emission direction of particles. Study of such dependence will yield more insight into medium modification of the away-side hadrons. Figure 5 shows the number and $p_T$-weighted correlation functions in central Au+Au collisions. The shaded areas show the correlated systematic uncertainties. The correlation functions are consistent with a flat distribution in $\Delta\phi$ within errors. Also shown are the minimum-bias pp and d+Au correlation functions (the systematic errors are not shown). The pp and d+Au data are similar. Both are peaked at $\Delta\phi = \pi$ and are significantly narrower than the central Au+Au data.

The $\langle p_T \rangle$ are obtained from the ratio of the background subtracted $p_T$-weighted over number correlation functions. Figure 6 shows the obtained $\langle p_T \rangle$ on the away side as a function of $\Delta\phi$ for $4 < p_T^{trig} < 6$ GeV/c and various collision systems in the upper panel and for central Au+Au collisions and various $p_T^{trig}$ windows in the lower panel. The inclusive hadron $\langle p_T \rangle$, obtained directly from the ratio of the subtracted backgrounds in Fig. 5 is shown as the straight horizontal lines (the small modulation due to $p_T$-dependent elliptic flow is invisible on this scale). The shaded areas depicted in both panels show the systematic uncertainties in central Au+Au collisions for $4<p_T^{trig}<6$ GeV/c. The $<p_T>$ is more robust than the correlation functions themselves because the background uncertainties in the number and $p_T$-weighted correlation functions are correlated and largely cancel in their ratio.

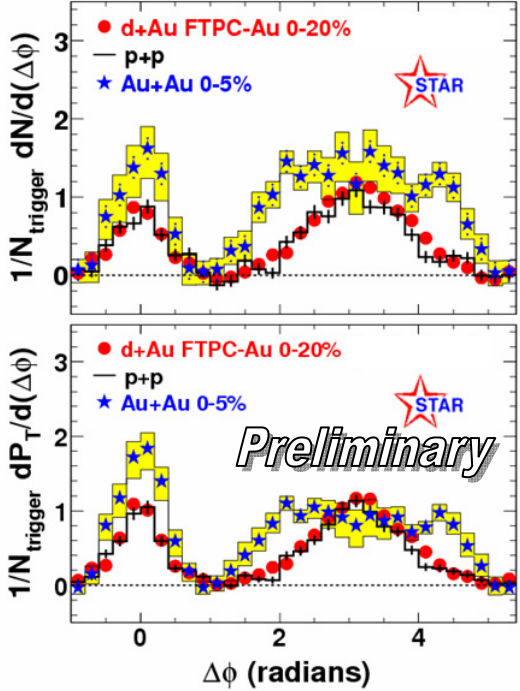 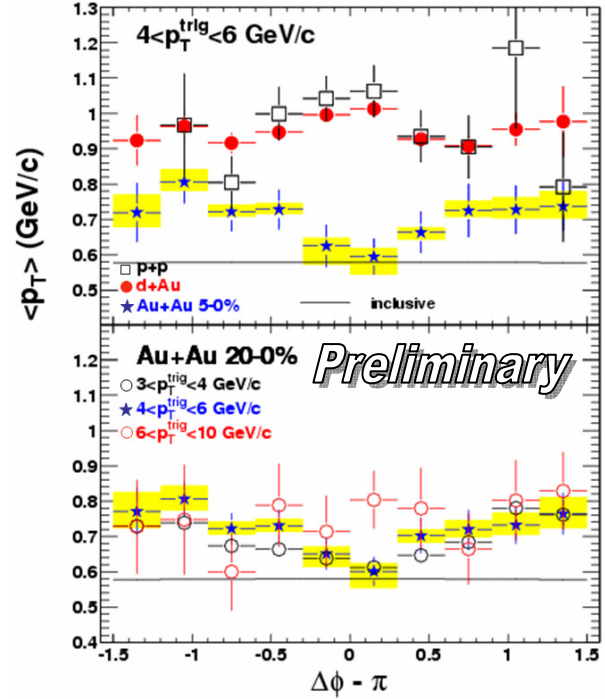

**Figure 5.** Background subtracted number (upper) and $p_T$-weighted (lower) correlation functions in min-bias pp, the 20% most central d+Au, and the 5% most central Au+Au collisions. The shaded areas are systematic uncertainties for the central Au+Au data.

**Figure 6.** The $<p_T>$ of associated hadrons on the away side. Upper panel shows those for min-bias pp, the 20% most central d+Au, and the 5% most central Au+Au collisions with trigger particle $4<p_T^{trig}<6$ GeV/c. Lower panel shows those for three trigger $p_T$ selections in the 20% most central Au+Au collisions. The shaded areas show the systematic uncertainties for central Au+Au collisions for the $4<p_T^{trig}<6$ GeV/c selection.

In Fig. 6 upper panel one observes that $<p_T>$ for pp and d+Au are peaked at $\Delta\phi = 180°$, and are much larger than that of inclusive hadrons. These features are expected from jet fragmentation: the fragment hadron momentum component perpendicular to the jet axis is independent of the parallel momentum component. Hence, fragments with larger parallel momentum component are more collimated with the jet axis. These hadrons are the relatively large $p_T$ particles in our measurements (which are made at mid-rapidity), resulting in a larger $<p_T>$ for $\Delta\phi$ closer to 180° from the trigger particle. For central Au+Au collisions, however, the $<p_T>$ is the smallest at $\Delta\phi = 180°$. The change from the peaked structure in pp, d+Au, and peripheral Au+Au (not shown) to the dipped structure in central Au+Au seems gradual with increasing centrality. The $<p_T>$ at $\Delta\phi = 180°$ in central Au+Au collisions appears equal to that of inclusive hadrons, while at other angles it is still larger. Figure 6 lower panel shows that the dipped shape of $<p_T>$ in central collisions is the same for $3<p_T^{trig}<4$ GeV/c and $4<p_T^{trig}<6$ GeV/c. The $<p_T>$ for $6<p_T^{trig}<10$ GeV/c is consistent with a constant distribution over $\Delta\phi$.

Our correlation function data are qualitatively consistent with the recently proposed sonic shock wave picture [28]. Sonic shock waves may result in a larger $<p_T>$ along the conical flow direction, approximately $\Delta\phi = \pi \pm 1$ [28]; However, theoretically this aspect is little explored. The seemingly flattening of the $<p_T>$ versus $\Delta\phi$ distribution from low to high trigger $p_T$ may be understood if one considers two contributions to the away-side associated hadrons: one from jet fragmentation and the other from sonic conical flow of the medium. Their relative contributions change with trigger $p_T$.

## 3. Deconfinement

The QGP is a state of deconfined quarks and gluons. While experimental search for deconfinement is elusive, recent theoretical and experimental developments in the constituent quark coalescence picture are encouraging [29].

3.1. Baryon/Meson Puzzle, Elliptic Flow Scaling, and Constituent Quark Coalescence

Recent results from PHENIX [30] and STAR [31,32] indicate that the high $p_T$ suppression is particle-type dependent. In central Au+Au collisions, charged and neutral pions [30] and $K_S$ [31,32] are suppressed by an approximately constant factor at $p_T > 3$ GeV/c while (anti-)protons [30] and (anti-)lambdas [31,32] are little suppressed in the intermediate $p_T$ range of $2 < p_T < 4$ GeV/c. The $p/\pi$ and $K_S/\Lambda$ ratios in central Au+Au collisions are 3 times that expected from jet fragmentation in the $2 < p_T < 4$ GeV/c range, start to drop at $p_T = 4$ GeV/c, and approach the corresponding jet fragmentation values at $p_T > 6$ GeV/c. The suppression pattern of K* [33] and $\phi$ [34] mesons seem to follow those of pions and kaons, rather than the similar mass protons and lambdas.

Moreover, the measured elliptic flow [35], well described by hydrodynamics at low $p_T$, saturates at $p_T > 2$ GeV. The saturation value depends on particle type: the $v_2$ of mesons is about 2/3 of the $v_2$ of baryons. This separation pattern holds for pions [36], kaons [31,36], K* [33], protons [36], lambdas [31], and cascades [37], and seems to hold for the $\Omega$ baryons [37].

These results suggest that neither the mass nor the species, but the baryon-meson difference, govern the dynamics of the intermediate $p_T$ region. Coalescence of constituent quarks [38] in the thermal bath at hadronization temperature gives an attractive explanation for both results: (1) It is more effective to produce baryons than mesons at the same $p_T$: To become a baryon, three quarks each at $p_T/3$ coalesce where they are much more abundant than at $p_T/2$ where two quarks coalesce into a meson, provided that the quark spatial density is so high that three quarks are equally available at same configuration space point as two quarks are. (2) The coalesced baryon will carry three times the elliptic flow of the constituent quarks while the coalesced meson, twice.

*If* constituent quark coalescence is indeed the production mechanism giving rise to the baryon-meson difference at intermediate $p_T$, i.e. hadronization of the bulk medium does occur, then it seems natural to conclude that a *deconfined* phase of quarks and gluons is created, prior to the hadronization. However, coalescence is expected to produce no jet-like angular correlations (in the case of coalescence of pure thermal partons), or weaker correlations than those from jet fragmentation (in the case of recombination of shower and thermal partons [39,40]). Therefore, jet-like angular correlations with a leading baryon and with a leading meson should generally differ in central Au+Au collisions as compared to pp.

3.2. Angular Correlations with Identified Trigger Particles

PHENIX [41] found little difference between angular correlations of hadrons in $1.7 < p_T < 2.5$ GeV/c with a proton and with a pion of $2.5 < p_T^{trig} < 4.0$ GeV/c. Figure 7 shows STAR preliminary results of $\Delta\phi$ correlations of charged hadrons in $0.15 < p_T < 3$ GeV/c with protons and with pions of $3 < p_T^{trig} < 4$ GeV/c. The protons and pions at high $p_T$ are identified by the relativistic rise of the specific ionization energy loss in the STAR TPC and are required to have 75% and 90% purity, respectively. The upper panel of Fig. 7 shows the correlation functions with a constant background subtraction. The lower panel has an additional subtraction of the elliptic flow effect. The $v_2$ values for trigger protons and pions are extracted from the measured charged hadron $v_2$ and the relative abundances of protons and pions, assuming 3/2 for the baryon/meson $v_2$ ratio. Little difference is found in Fig. 7 between the $\Delta\phi$ correlation functions for trigger protons and trigger pions, and additionally between particles and anti-particles. STAR has also measured angular correlations with leading (anti-)lambdas and leading $K_S$, and also little difference is found [42].

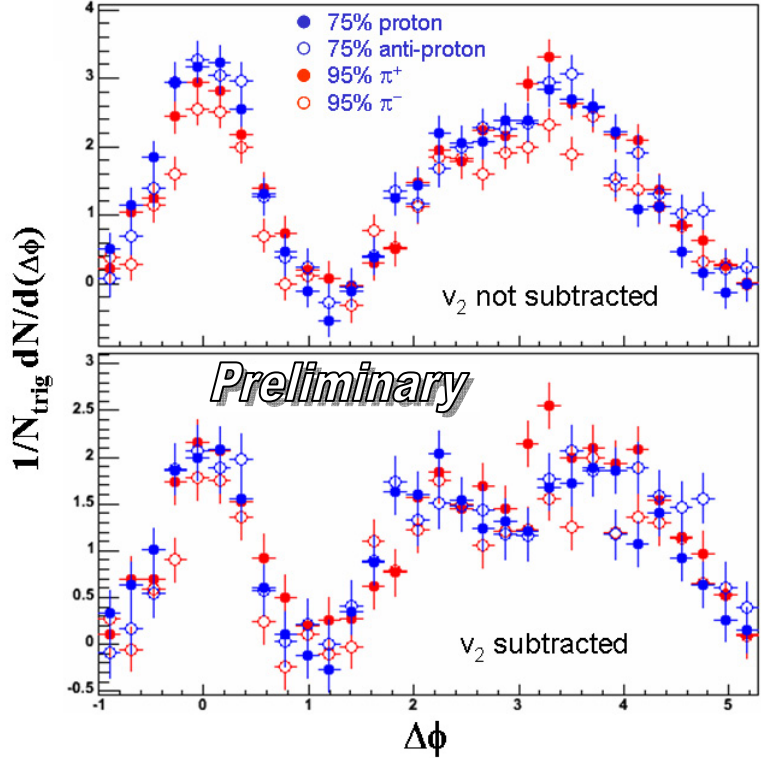

Figure 7. Azimuthal correlation functions of charged hadrons in the range of $0.15 < p_T < 3$ GeV/c with identified trigger particles (proton, anti-proton, $\pi^+$, and $\pi^-$) of $3 < p_T < 4$ GeV/c. Only a constant background is subtracted in the upper panel; The full background with elliptic flow modulation is subtracted in the lower panel.

Clearly, the success of the coalescence/recombination model in describing the single particle yield and elliptic flow measurements, but its apparent difficulty in explaining the observed indifference between jet-like correlations with leading baryons and leading mesons, needs to be reconciled. It will benefit from more detailed measurements, with greater statistics, of angular correlations with identified particles. While it is still tempting to conclude that deconfined quarks and gluons are likely, a firmer conclusion is premature.

**4. Summary**
The measured suppression magnitudes of high $p_T$ yields and jet-like angular correlations in central Au+Au collisions at RHIC suggest, within the framework of pQCD models with partonic energy loss, an initial medium energy density two orders of magnitude larger than normal nuclear density, well above the critical energy density for QGP formation. The reconstructed correlated hadrons over a wide $p_T$ range on the away side of a triggered jet are broadly distributed in azimuthal angle and become partially thermalized with the bulk medium, suggesting a high degree of thermalization in the bulk medium itself. The division between baryon and meson results in the intermediate $p_T$ range suggests the intriguing relevance of the constituent quark degrees of freedom at hadronization, which may be the first experimental hint for deconfinement.

The experimental evidences for QGP formation seem strong. A firmer conclusion, however, requires further systematic investigations, with high statistics data at several bombarding energies and with several colliding nuclei species, to verify that the existing models and interpretations indeed give a consistent description of the dynamics in heavy-ion collisions.